\documentclass[twocolumn,aps,prl,superscriptaddress,amsfonts]{revtex4-1}

\usepackage{amsmath}
\usepackage{graphicx}
\usepackage{bm}
\usepackage{array}
\usepackage{dcolumn}
\usepackage{hepparticles}
\usepackage{heppennames}
\usepackage{hepnicenames}
\usepackage{epstopdf}

\begin{document}

\def\e{\epsilon}
\def\d{\downarrow}
\def\u{\uparrow}

\def\e{\mathcal{E}}

\def\ba{\begin{eqnarray}}
\def\ea{\end{eqnarray}}
\def\beq{\begin{equation}}
\def\eeq{\end{equation}}

\newcommand{\ket}[1]{{| {#1} \rangle}}
\newcommand{\bra}[1]{{\langle {#1} |}}

\newcommand{\FigureOne}{
\begin{figure}[b!]
\includegraphics*[width=\columnwidth]{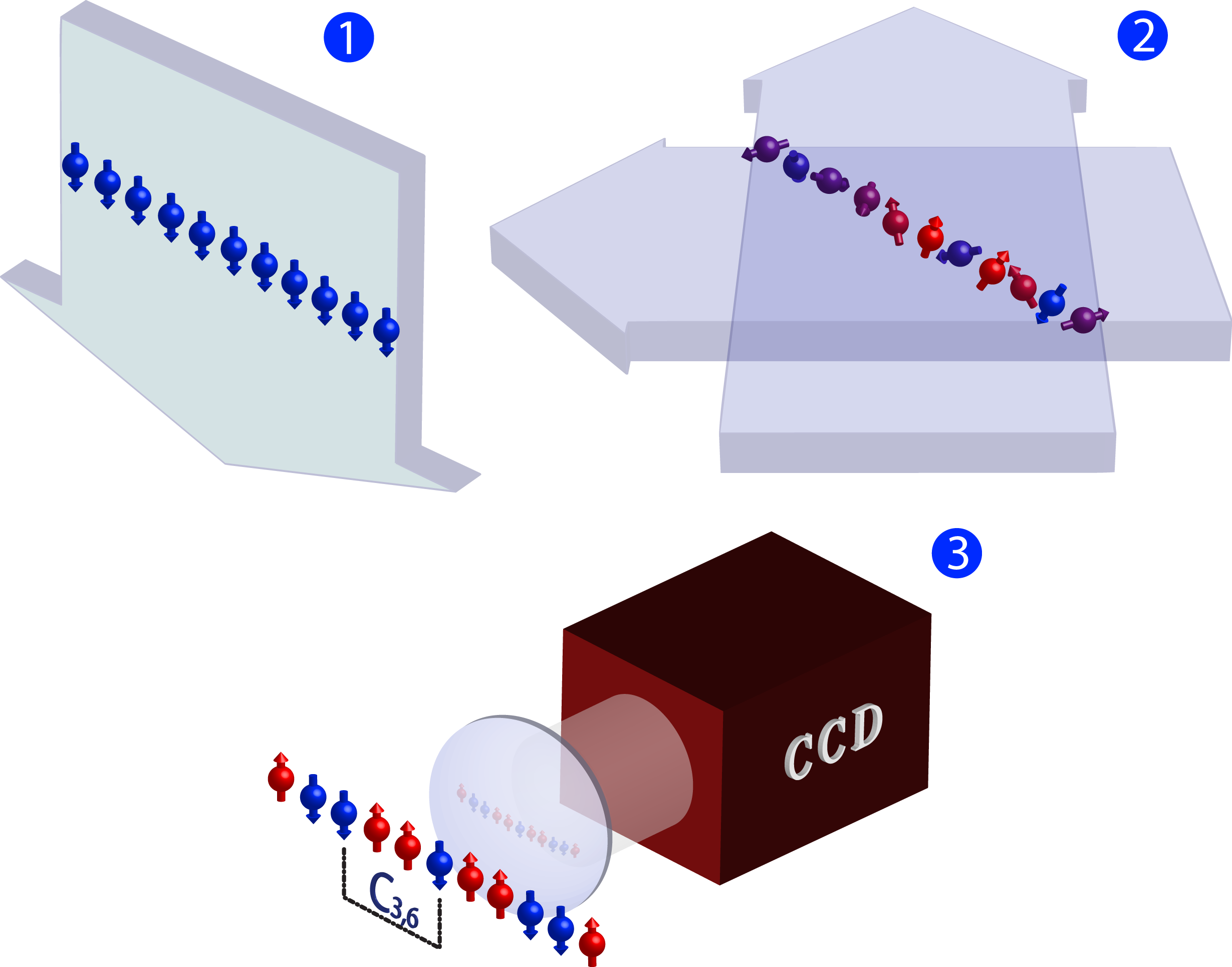}
\caption{(1) The experiment is initialized by optically pumping all 11 spins to the state $\ket{\downarrow}_z$. (2) After initialization, the system is quenched by applying laser-induced forces on the ions, yielding an effective Ising or $XY$ spin chain (see text for details). After allowing dynamical evolution of the system, the projection of each spin along the $\hat{z}$ direction is imaged onto a CCD camera (3). Such measurements allow us to construct any possible correlation function $C_{i,j}$ along $\hat{z}$.}
\label{fig:Cartoon}
\end{figure}
}

\newcommand{\FigureTwo}{
\begin{figure*}[t]
\includegraphics[width=\textwidth]{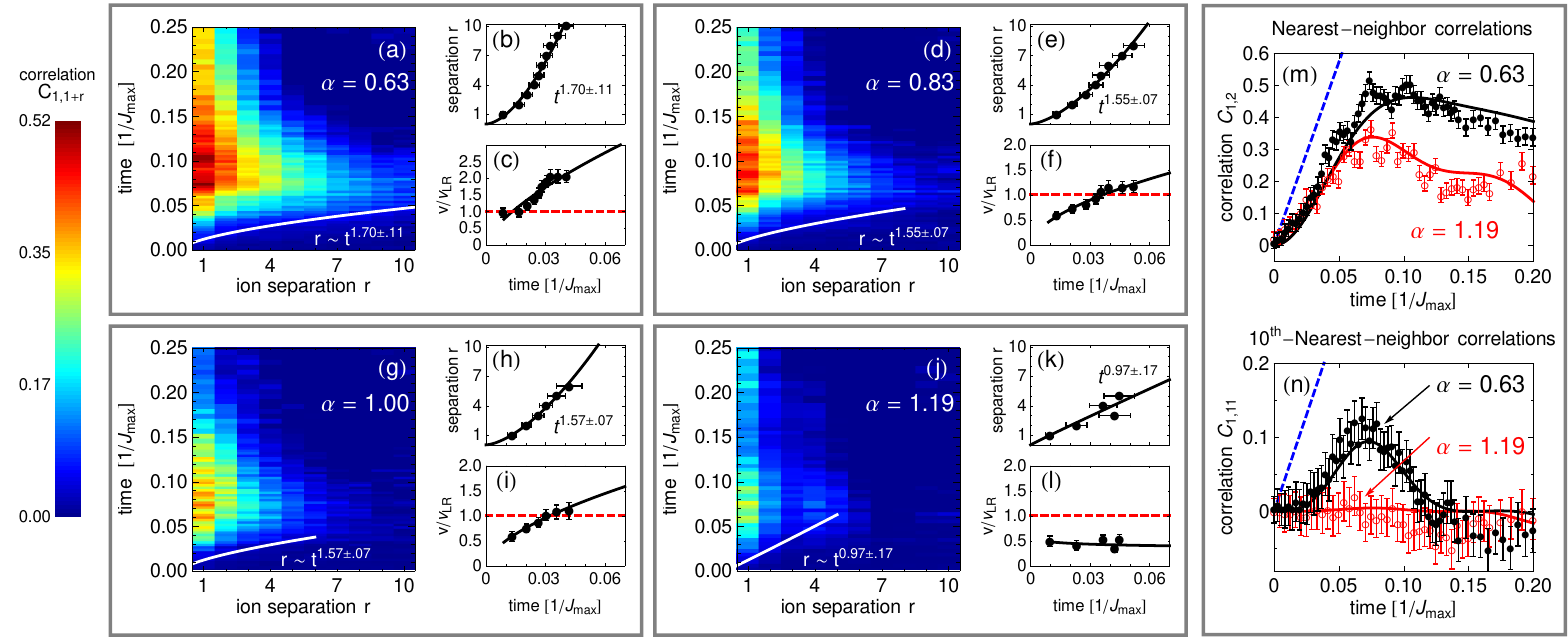}
\caption{(a-c): Spatial and time-dependent correlations (a), extracted light-cone boundary (b), and correlation propagation velocity (c) following a global quench of a long-range Ising model with $\alpha=0.63$. The curvature of the boundary shows an increasing propagation velocity (b), quickly exceeding the short-range Lieb-Robinson velocity bound, $v_{\text{LR}}$ (c). Solid lines give a power-law fit to the data, which slightly depends on the choice of fixed contour $C_{i,j}$. Complementary plots are shown for $\alpha=0.83$ (d-f), $\alpha=1.00$ (g-i), and $\alpha=1.19$ (j-l). As the system becomes shorter-range, correlations do not propagate as far or as quickly through the chain; the short-range velocity bound $v_{\text{LR}}$ is not exceeded for our shortest-range interaction. (m,n): Nearest- and 10$^\text{th}$-nearest-neighbor correlations for our shortest- and longest-range interaction compared to the exact solution (i.e. no free parameters) from Eq.\ \ref{eqn:ExactIsing} (solid). The dashed blue curves show a long-range bound for any commuting Hamiltonian (see Supplementary Information).}
\label{fig:Ising}
\end{figure*}
}

\newcommand{\FigureThree}{
\begin{figure*}[t!]
\includegraphics*[width=\textwidth]{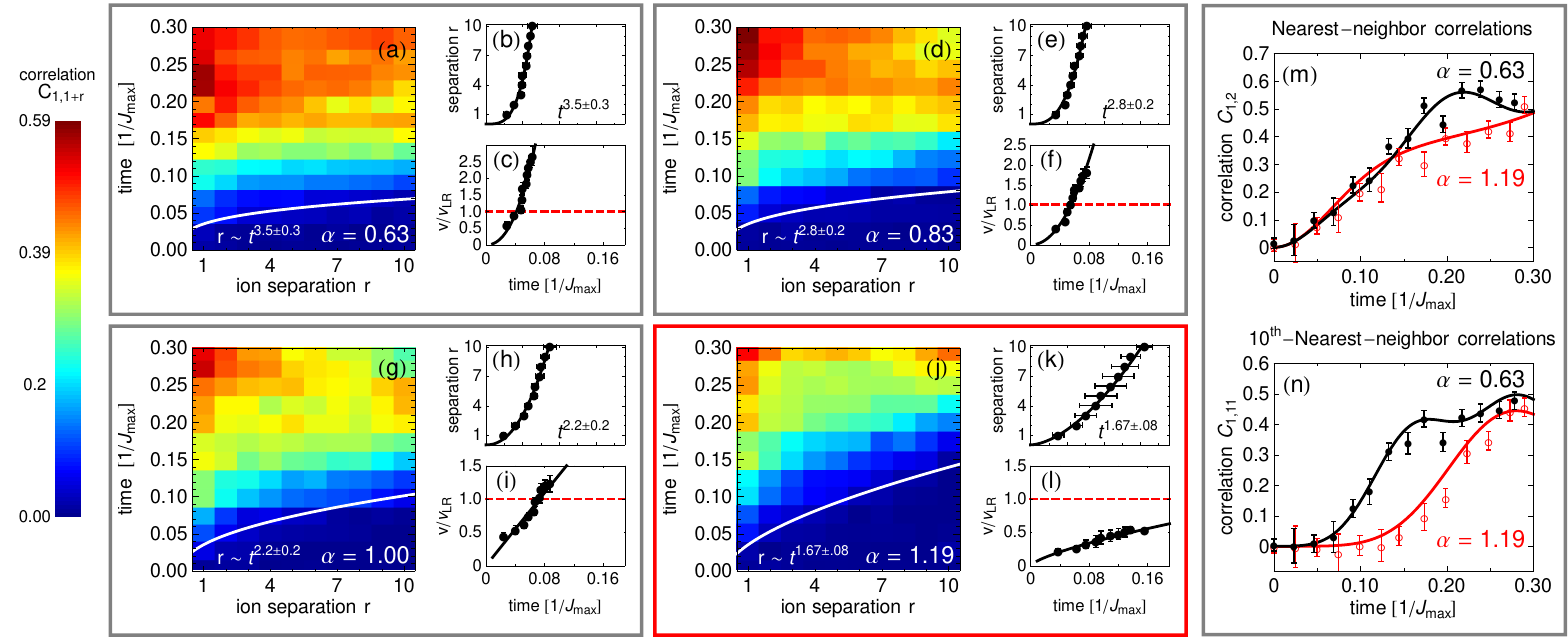}
\caption{Global quench of a long-range $XY$ model with four different interaction ranges. (a-l): Panel descriptions match those in Fig.\ \ref{fig:Ising}. In each case, when compared with the Ising model, correlations between distant sites in the $XY$ model are stronger and build up more quickly. For the shortest-range interaction [red box, (j-l)], we observe a faster-than-linear growth of the light-cone boundary, despite $\alpha > 1$; no known analytic theory predicts this effect. (m,n): Nearest- and 10$^\text{th}$-nearest-neighbor correlations compared to a solution found by numerically evolving the Schr\"odinger equation of an $XY$ model with experimental spin-spin couplings.}
\label{fig:XY}
\end{figure*}
}

\newcommand{\FigureFour}{
\begin{figure}[t!]
\includegraphics*[width=\columnwidth]{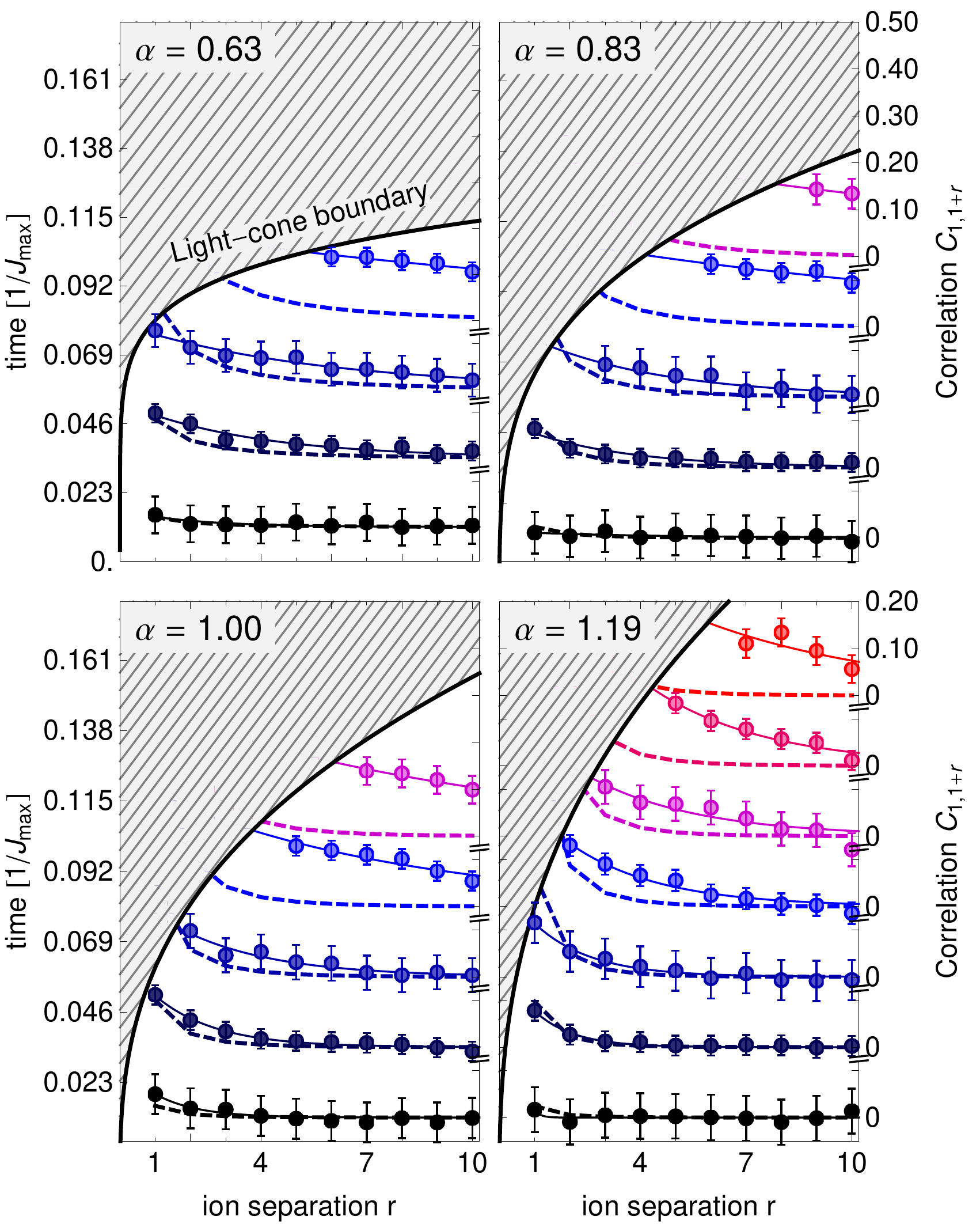}
\caption{Decay of spatial correlations outside the light-cone boundaries for a long-range $XY$ model with $\alpha=\{0.63, 0.83, 1.00, 1.19\}$. The hatched region indicates the area inside the light-cone boundary $C_{i,j}=0.15$. The data corresponds to times indicated by tickmarks on the left axis. Solid lines give an exponential fit to the data while dashed lines show the predictions from a perturbative calculation. Perturbation theory does not accurately describe the dynamics at later times.}
\label{fig:XYdecay}
\end{figure}
}

\newcommand{\SuppFigureOne}{
\begin{figure}[h!t]
\includegraphics*[width=\columnwidth]{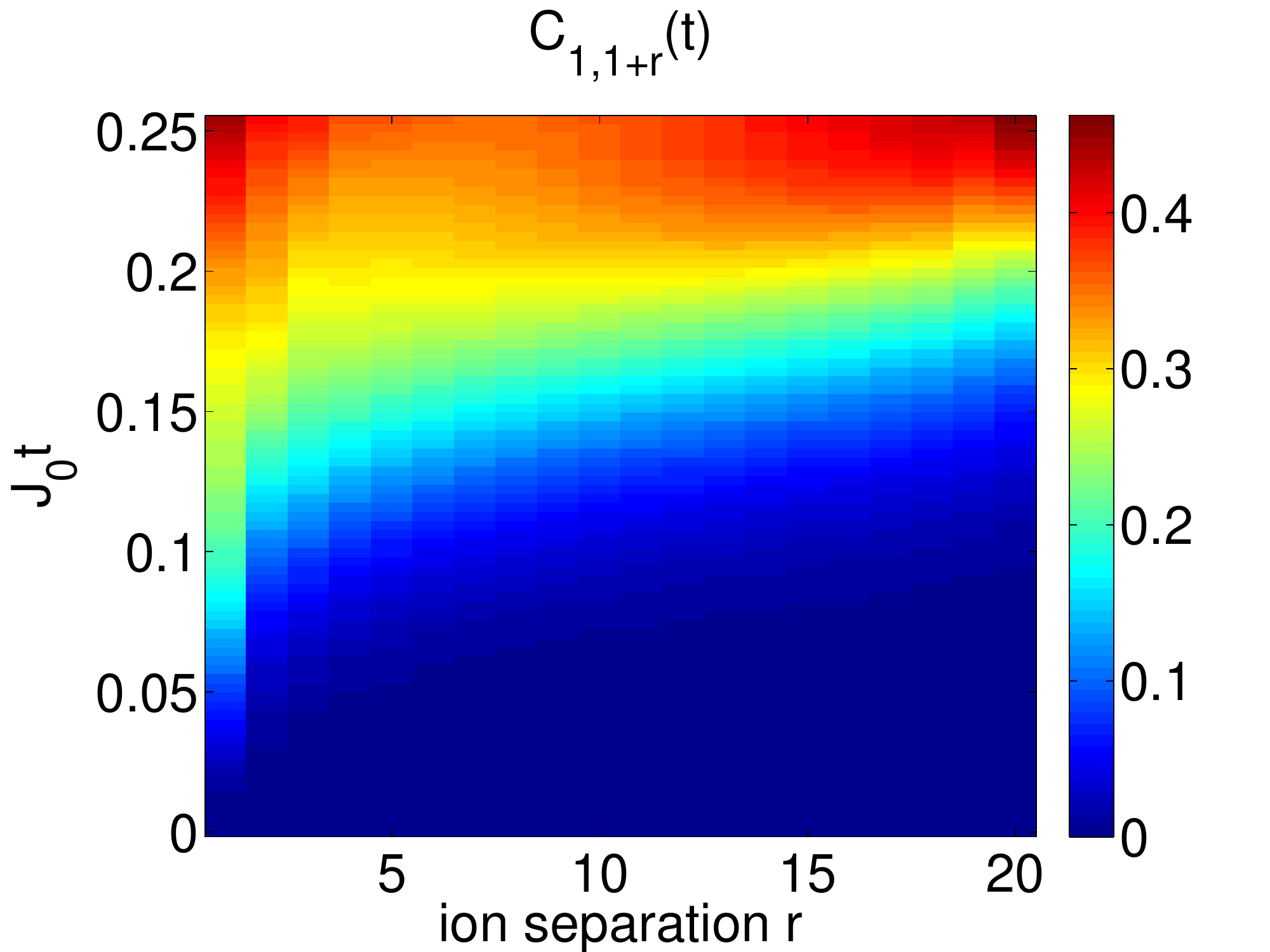}
\caption{Calculated spatial and time-dependent correlations for an $XY$ model with spin-spin couplings \mbox{$J_{ij}=J_0/|i-j|^{1.19}$}, found by numerically evolving the Schr\"odinger equation.}
\label{fig:LargeSimulation}
\end{figure}
}

\title{Non-local propagation of correlations in long-range interacting quantum systems}

\author{\mbox{P. Richerme$^1$, Z.-X. Gong$^1$, A. Lee$^1$, C. Senko$^1$, J. Smith$^1$, M. Foss-Feig}}
\affiliation{Joint Quantum Institute, University of Maryland Department of Physics and National Institute of Standards and Technology, College Park, MD 20742}

\author{S. Michalakis}
\affiliation{Institute for Quantum Information and Matter, California Institute of Technology, Pasadena, CA 91125}

\author{A. V. Gorshkov}
\affiliation{Joint Quantum Institute, University of Maryland Department of Physics and National Institute of Standards and Technology, College Park, MD 20742}

\author{C. Monroe}
\affiliation{Joint Quantum Institute, University of Maryland Department of Physics and National Institute of Standards and Technology, College Park, MD 20742}

\date{\today}

\begin{abstract}
The maximum speed with which information can propagate in a quantum many-body system directly affects how quickly disparate parts of the system can become correlated \cite{Nachtergaele2006,Schachenmayer2013,HaukeTagliacozzo,KadenPRL2013} and how difficult the system will be to describe numerically \cite{AreaLawReview}. For systems with only short-range interactions, Lieb and Robinson derived a constant-velocity bound that limits correlations to within a linear effective light cone \cite{LiebRobinson}. However, little is known about the propagation speed in systems with long-range interactions, since the best long-range bound \cite{HastingsKoma} is too loose to give the correct light-cone shape for any known spin model and since analytic solutions rarely exist. In this work, we experimentally determine the spatial and time-dependent correlations of a far-from-equilibrium quantum many-body system evolving under a long-range Ising- or XY-model Hamiltonian. For several different interaction ranges, we extract the shape of the light cone and measure the velocity with which correlations propagate through the system. In many cases we find increasing propagation velocities, which violate the Lieb-Robinson prediction, and in one instance cannot be explained by any existing theory. Our results demonstrate that even modestly-sized quantum simulators are well-poised for studying complicated many-body systems that are intractable to classical computation.
\end{abstract}

\maketitle

Lieb-Robinson bounds \cite{LiebRobinson} have strongly influenced our understanding of locally-interacting quantum many-body systems. These bounds restrict the many-body dynamics to a well-defined causal region outside of which correlations are exponentially suppressed \cite{BravyiHastingsVerstraete}, analogous to causal light cones that arise in relativistic theories. Their existence has enabled proofs linking the decay of correlations in ground states to the presence of a spectral gap \cite{HastingsKoma,LiebRobinsonExponentialClustering}, as well as the area law for entanglement entropy \cite{HastingsAreaLaw,SpirosAreaLaw,AreaLawReview}, which can indicate the computational complexity of classically simulating a quantum system. Furthermore, Lieb-Robinson bounds constrain the timescales on which quantum systems might thermalize \cite{RigolThermalizationPRL,CalabreseCardy2006,ZhexuanPrethermalization} and the maximum speed with which information can be sent through a quantum channel \cite{BoseSpinChainReview}. Recent experimental work has observed an effective Lieb-Robinson (i.e. linear) light cone in a 1D quantum gas \cite{BlochLightCone}.

When interactions in a quantum system are long-range, the speed with which correlations build up between distant particles is no longer guaranteed to obey the Lieb-Robinson prediction. Indeed, for sufficiently long-ranged interactions, the notion of locality is expected to break down completely \cite{EisertLightCone}. Violation of the Lieb-Robinson bound means that comparatively little can be predicted about the growth and propagation of correlations in long-range interacting systems, though there have been several recent theoretical and numerical advances \cite{Schachenmayer2013,HaukeTagliacozzo,HastingsKoma,ExactIsingCorrelations,EisertLightCone,JQITheoryBlank}.

\FigureOne
\FigureTwo

Here we report an experiment that directly measures the shape of the causal region and the speed at which correlations propagate within Ising and $XY$ spin chains. To induce the spread of correlations, we perform a global quench by suddenly switching on the spin-spin couplings across the entire chain and allowing the system to coherently evolve. The dynamics following a global quench can be highly non-intuitive; one picture is that entangled quasi-particles created at each site propagate outwards, correlating distant parts of the system through multiple interference pathways \cite{CalabreseCardy2006}. This process differs substantially from local quenches, where a single site emits quasi-particles that may travel ballistically \cite{CalabreseCardy2006,HaukeTagliacozzo}, resulting in a different causal region and propagation speed than in a global quench (even for the same spin model). An experimental study of local quenches appears in \cite{LanyonBlank}.

In our experiment, the effective spin-1/2 system is encoded into the $^2$S$_{1/2}\ket{F=0,m_F=0}$ and $\ket{F=1,m_F=0}$ hyperfine `clock' states of trapped atomic $^{171}$Yb$^+$ ions, denoted $\ket{\downarrow}_z$ and $\ket{\uparrow}_z$, respectively \cite{YbDetection}. The ions are confined in a 3-layer rf Paul trap with a \mbox{$4.8$ MHz} radial frequency and form a linear chain along the central axis. Long-range interactions are mediated by phonons which couple the ions through their collective modes of motion \cite{PorrasCiracQSIM,QSIMNature2010,QSIMAxialField,QSIMPRL2009,QSIM2013Science}.

We initialize a chain of 11 ions by optically pumping to the product state $\ket{\downarrow\downarrow\downarrow\ldots}_z$ (see Fig.\ \ref{fig:Cartoon}). At $t = 0$, we quench the system by applying laser-induced optical dipole forces \cite{Molmer1999,PorrasCiracQSIM,QSIMPRL2009} to yield an Ising-model Hamiltonian

\begin{equation}
\label{eqn:Ising}
H_\text{Ising}=\sum_{i<j} J_{i,j}\sigma_i^x\sigma_j^x
\end{equation}
or an $XY$-model Hamiltonian
\begin{equation}
\label{eqn:XY}
H_{XY}=\frac{1}{2}\sum_{i<j} J_{i,j}(\sigma_i^x\sigma_j^x+\sigma_i^z\sigma_j^z),
\end{equation}
where $\sigma_i^\gamma$ ($\gamma=x,y,z$) is the Pauli spin matrix acting on the $i^\text{th}$ spin, $h=1$, and $J_{i,j}$ (in cyclic frequency) gives the coupling strength between spins $i$ and $j$. 

For both model Hamiltonians, the spin-spin interaction matrix $J_{i,j}$ contains tunable, long-range couplings that fall off approximately algebraically as $J_{i,j} \propto 1/|i-j|^\alpha$. We vary the interaction range $\alpha$ by adjusting a combination of trap and laser parameters \cite{QSIM2013Science} (see Methods), choosing $\alpha\approx\{0.63, 0.83, 1.00, 1.19\}$ for these experiments. For values $\alpha <1$, the system is strongly long-range, meaning that in the thermodynamic limit the interaction energy per site diverges, and so the generalized Lieb-Robinson bound of Ref. \cite{HastingsKoma} breaks down.

After quenching to the Ising or $XY$ model with our chosen value of $\alpha$, we allow coherent evolution for various lengths of time before resolving the spin state of each ion using a CCD camera. The experiments at each time step are repeated 4000 times to collect statistics. To observe the buildup of correlations, we use the measured spin states to construct the connected correlation function
\begin{equation}
\label{eqn:Cij}
C_{i,j}(t) = \langle \sigma_i^z(t)\sigma_j^z(t)\rangle-\langle \sigma_i^z(t)\rangle\langle \sigma_j^z(t)\rangle
\end{equation}
between any pair of ions at any time. Since the system is initially in a product state, $C_{i,j}(0)=0$ everywhere. As the system evolves away from a product state, evaluating Eqn. \ref{eqn:Cij} at all points in space and time provides the shape of the light-cone boundary and the correlation propagation velocity for our long-range spin models.
\FigureThree

\bigskip
\textbf{Ising Model -- } Figure \ref{fig:Ising} shows the results of globally quenching the system to a long-range Ising model for four different interaction ranges. In each case, we extract the light-cone boundary by observing the time it takes a correlation of fixed amplitude (here, $C_{i,j}=0.04 \approx 0.1 C_{i,j}^{\text{max}}$) to travel an ion-ion separation distance $r$. For strongly long-range interactions ($\alpha < 1$), the region within the light-cone grows faster than linearly, which violates the Lieb-Robinson prediction. This fast propagation of correlations is not surprising, because even the direct long-range coupling between distant spins produces correlations in a time $t \propto 1/J_{i,j} \sim r^\alpha$. Thus, faster-than-linear light-cone shapes are expected to be a general feature of any 1D long-range interacting Hamiltonian with $\alpha < 1$. 

Increasing propagation velocities quickly surpass the Lieb-Robinson velocity for a system with nearest-neighbor interactions, $v_{LR}=12eJ_{\text{max}}$ [see Fig.\ \ref{fig:Ising}(c,f,i)]. Such violations indicate that predictions based on the Lieb-Robinson result -- including those that bound the growth of entanglement or correlation lengths in the system -- can no longer be trusted. However, for the specific case of the pure Ising model, the correlations at any time can be predicted by an exact analytic solution \cite{ExactIsingCorrelations,MikeIsingPRA}:
\begin{eqnarray}
\label{eqn:ExactIsing}
\nonumber
C_{i,j}(t)&=&\frac{1}{2}\prod_{k\neq i,j} \cos[2(J_{i,k}+J_{j,k})t]\\
&+&\frac{1}{2}\prod_{k\neq i,j} \cos[2(J_{i,k}-J_{j,k})t]\\
\nonumber
&-&\prod_{k\neq i}\cos[2 J_{i,k}t]\prod_{k\neq j} \cos[2 J_{j,k} t].
\end{eqnarray}

In Eq.\ \ref{eqn:ExactIsing}, correlations can only build up between sites $i$ and $j$ that are coupled either directly or through a single intermediate spin $k$; processes which couple through more than one intermediate site are prohibited. For instance, if the $J_{i,j}$ couplings are nearest-neighbor-only, $C_{i,j}(t)=0$ for all $|i-j| > 2$. This property holds for any commuting Hamiltonian (see Supplementary Information) and explains why the spatial correlations shown in Fig.\ \ref{fig:Ising} become weaker for shorter-range systems.

The products of cosines in Eq.\ \ref{eqn:ExactIsing} with many different oscillation frequencies result in the observed decay of correlations when $t \gtrsim 0.1/J_{\text{max}}$. At later times, rephasing of these oscillations creates revivals in the spin-spin correlation. One such partial revival occurs at $t=2.44/J_{\text{max}}$ for the $\alpha=0.63$ case (not shown), verifying that our system remains coherent for a timescale much longer than that which determines the light-cone boundary. 

\bigskip
\textbf{$XY$ Model -- }We repeat the quench experiments for an $XY$-model Hamiltonian using the same set of interaction ranges $\alpha$, as shown in Fig.\ \ref{fig:XY}. Dynamical evolution and the spread of correlations in long-range interacting $XY$ models are much more complex than in the Ising case because the Hamiltonian contains non-commuting terms. As a result, no exact analytic solution comparable to \mbox{Eq.\ \ref{eqn:ExactIsing}} exists.

Compared with the correlations observed for the Ising Hamiltonian, correlations in the $XY$ model are much stronger at longer distances [e.g. Fig.\ \ref{fig:Ising}(j) vs. Fig.\ \ref{fig:XY}(j)]. Processes coupling through multiple intermediate sites (which were disallowed in the commuting Ising Hamiltonian) now play a critical role in building correlations between distant spins. These processes may also explain our observation of a steeper power-law scaling of the light-cone boundary in the $XY$ model. However, we note that without an exact solution, there is no \emph{a priori} reason to assume a power-law light-cone edge (used for the fits in Fig.\ \ref{fig:XY}), and deviations from power-law behavior might reveal themselves for larger system sizes.

An important observation in Fig.\ \ref{fig:XY}(j-l) is that of faster-than-linear light-cone growth for the relatively short-range interaction $\alpha=1.19$. Although faster-than-linear growth is expected for $\alpha < 1$ (see previous section), there is no consensus on whether such behavior is generically expected for $\alpha > 1$. Our experimental observation has prompted us to numerically check the light-cone shape for $\alpha=1.19$; we find that faster-than-linear scaling persists in systems of up to 22 spins before our calculations break down. Whether such scaling continues beyond $\sim 30$ spins is a question that, at present, quantum simulators are best positioned to answer. 

For the $XY$ model, we additionally study the spatial decay of correlations outside the light-cone boundary. The data is shown in Fig.\ \ref{fig:XYdecay} and is well-described by fits to exponentially decaying functions. Recent theoretical work \cite{JQITheoryBlank} predicts an initial decay of spatial correlations bounded by an exponential, followed by a power-law decay; we speculate that much larger system sizes and several hundred-thousand repetitions of each data point (to sufficiently reduce the shot-noise uncertainty) would be necessary to see this effect. 

A perturbative treatment of time evolution under the $XY$ Hamiltonian yields the short-time approximation for the correlation function $C_{i,j}(t) \approx (J_{i,j} t)^2$. These values are plotted as dashed lines along with the data in Fig.\ \ref{fig:XYdecay}. While the perturbative result matches the data early on, it clearly fails to describe the dynamics at longer evolution times. The discrepancies indicate that the light-cone shapes observed in the $XY$ model are fundamentally non-perturbative; rather, they result from the build-up of correlations through multiple intermediate sites and cannot be understood by any known analytical method.

\FigureFour

We have presented experimental observations of the causal region and propagation velocities for correlations following global quenches in Ising and $XY$ spin models. The long-range interactions in our system lead to a breakdown of the locality associated with Lieb-Robinson bounds, while dynamical evolution in the $XY$ model leads to results that cannot be described by analytic or perturbative theory. Our work demonstrates that even modestly-sized quantum simulators can be an important tool for investigating and enriching our understanding of dynamics in complex many-body systems.

\bibliographystyle{prsty}
\bibliography{qsimrefs}

\section{Methods}
\textbf{Ising and $XY$ Couplings}
Spin-spin interactions are generated by applying spin-dependent optical dipole forces to the trapped ion chain. Two off-resonant laser beams with a wavevector difference $\Delta k$ along a principal axis of transverse motion globally address the ions and drive stimulated Raman transitions. The two beams contain a pair of beatnote frequencies that are symmetrically detuned from the resonant transition at $\nu_0=12.642819$ GHz by a frequency $\mu$ that is comparable to the transverse motional mode frequencies. In the Lamb-Dicke regime \cite{LeibfriedIonTrapReview}, this results in the Ising-type Hamiltonian in Eq.\ \ref{eqn:Ising} \cite{Molmer1999,PorrasCiracQSIM} with spin-spin coupling strengths 
\begin{equation}
\label{eqn:Jij}
J_{i,j}=\Omega^2\omega_R \sum_{m=1}^N \frac{b_{i,m}b_{j,m}}{\mu^2-\omega_m^2},
\end{equation}
where $\Omega$ is the global Rabi frequency at each ion, \mbox{$\omega_R=\hbar\Delta k^2/(2M)$} is the recoil frequency, $b_{i,m}$ is the normal-mode matrix \cite{James1998}, and $\omega_m$ are the transverse mode frequencies. The coupling profile may be approximated as a power-law decay $J_{i,j}\approx J_0/|i-j|^\alpha$, where in principle $\alpha$ can be tuned between 0 and 3 by varying the laser detuning $\mu$ or the trap frequencies $\omega_m$.

An effective transverse magnetic field $B\sum_i \sigma_i^y$ can be added to the pure Ising Hamiltonian by applying an additional laser beatnote frequency at $\nu_0$ that drives Rabi oscillations. In the limit $B \gg J$, processes in the $\sigma_i^x\sigma_j^x$ coupling which flip two spins along $y$ (e.g. $\sigma^+\sigma^+$, where here $\sigma^{\pm}=\sigma^z\pm i\sigma^x$) are energetically forbidden, leaving only the energy conserving flip-flop terms ($\sigma^+\sigma^- + \sigma^-\sigma^+$). At times $t=n/B$ (with integer $n$), the dynamics of the transverse field rephase and leave only the pure $XY$ Hamiltonian of \mbox{Eq.\ \ref{eqn:XY}.}

In the limit $B > \eta_m\Omega$, where $\eta_m=\Delta k \sqrt{\hbar/(2M\omega_m)}$, phonon contributions from the large transverse field can lead to unwanted spin-motion entanglement at the end of an experiment. Therefore, this method of generating an $XY$ model requires the hierarchy $J \ll B \ll \eta_m\Omega$ for all $m$. For our typical trap parameters, $J_{\text{max}}\approx 400$ Hz, $B\approx 4$ kHz, and $\eta_m\Omega\approx$ 20 kHz.

\section{Acknowledgements}
We thank J. Preskill, A.M. Rey, K. Hazzard, A. Daley, J. Schachenmayer, M. Kastner, S. Manmana, and L.-M. Duan for helpful discussions. This work is supported by the U.S. Army Research Office (ARO) Award W911NF0710576 with funds from the DARPA Optical Lattice Emulator Program, ARO award W911NF0410234 with funds from the IARPA MQCO Program, and the NSF Physics Frontier Center at JQI. MFF thanks the NRC for support.

\appendix*
\renewcommand{\theequation}{S\arabic{equation}}
\renewcommand{\thefigure}{S\arabic{figure}}

\section{SUPPLEMENTARY INFORMATION}
\subsection{State detection and readout}
After quenching to and allowing time evolution under our spin Hamiltonian, we measure the spin projections of each ion along the $z$ direction of the Bloch sphere. For \mbox{3 ms}, we expose the ions to a laser beam that addresses the cycling transition $^2$S$_{1/2}\ket{F=1}$ to $^2$P$_{1/2}\ket{F=0}$. Ions fluoresce only if they are in the state $\ket{\uparrow}_z$. This fluorescence is collected through an NA=0.23 objective and imaged using an intensified CCD camera with single-site resolution.

To discriminate between `bright' and `dark' states  ($\ket{\uparrow}_z$ and $\ket{\downarrow}_z$, respectively), we begin by calibrating the camera with 1000 cycles each of all-bright and all-dark states. For the bright states, the projection of the 2D CCD image onto a one-dimensional row gives a profile comprised of Gaussians at each ion location. We perform fits to locate the center and fluorescence width of each ion on our CCD.

We achieve single-shot discrimination of individual ion states in the
experimental data by fitting the captured one-dimensional profile to a
series of Gaussians with calibrated widths and positions but
freely-varying amplitudes. The extracted amplitudes for each ion are
then compared with a threshold found via Monte-Carlo simulation to
determine whether the measured state was `bright' or `dark'. Our
discrimination protocol also gives an estimate of the detection error
(e.g. misdiagnosing a `bright' ion as `dark'), which is typically of
order $\sim 5\%$. Corrected state probabilities (along with their
respective errors) are found following the method outlined in Ref. \cite{CorrectingDetectionErrors}, which also takes into account errors due to quantum projection noise.

\subsection{Lieb-Robinson velocity for nearest-neighbor interactions}
Here we justify our claim that the Lieb-Robinson velocity
\cite{LiebRobinson} for the spread of correlation functions from an
initial product state, evolving under  a 1D
spin Hamiltonian with only nearest-neighbor interactions, is bounded
above by
$v_{\rm LR}=12eJ$.  In particular, we consider a Hamiltonian
\begin{equation}
H=\sum_{j}h_{j,j+1},
\end{equation}
with interaction strength $\lVert h_{j,j+1}\rVert=J$.  Note that both the Ising and $XY$ Hamiltonians defined in the
manuscript satisfy these assumptions in the $\alpha\rightarrow\infty$ limit, where $J_{ij}=J\delta_{j,i+1}$, as can easily be checked by
calculating
$\lVert\sigma^{x}_i\sigma^x_j\rVert=\lVert\sigma^{x}_i\sigma^x_j+\sigma^{z}_i\sigma^z_j\rVert/2=1$.
For operators evolving in the Heisenberg picture under $H$,
$A(t)\equiv e^{iHt}A(0)e^{-iHt}$, we would like to compute the connected correlation function
\begin{equation}
C_{i,j}(t)=\langle A_{i}(t)B_{j}(t)\rangle_c\equiv \langle
A_{i}(t)B_{j}(t)\rangle-\langle A_{i}(t)\rangle\langle B_{j}(t)\rangle,
\end{equation}
where $A_i$ and $B_j$ are supported on sites $i$ and $j$,
respectively.

A bound on these correlation functions follows immediately from
results in Ref.~\cite{BravyiHastingsVerstraete}, which relate a Lieb-Robinson bound on unequal-time commutators to a bound on connected correlation functions.  In
particular, for a Lieb-Robinson commutator bound of the form
\begin{equation}
\lVert\left[A_{i}(t),B_{j}(0)\right]\rVert\leq c\lVert
A_{i}\rVert\lVert
B_{j}\rVert e^{(vt-r)/\xi},
\end{equation}
we have
\begin{equation}
\label{eq:cijbound}
C_{i,j}(t)\leq 4c\lVert
A_{i}\rVert\lVert
B_{j}\rVert e^{(vt-r/2)/\xi},
\end{equation}
where $r$ is the distance between the two sites $i$ and $j$.

The Lieb-Robinson commutator bound for a nearest-neighbor Hamiltonian on a
D-dimensional square lattice is
given by (Ref. \cite{BravyiHastingsVerstraete})
\begin{equation}
\lVert\left[A_{i}(t),B_{j}(0)\right]\rVert\leq 2\lVert
A_{i}\rVert\lVert
B_{j}\rVert\sum_{k=r}^{\infty}\frac{(2Jt(4D-1))^k}{k!},
\end{equation}
which in 1D gives
\ba
\lVert\left[A_{i}(t),B_{j}(0)\right]\rVert&\leq&2\lVert
A_{i}\rVert\lVert
B_{j}\rVert e^{-r}\sum_{k=r}^{\infty}\frac{(6eJt)^k}{k!}\\
&\leq& 2\lVert
A_{i}\rVert\lVert
B_{j}\rVert e^{6eJt-r},
\ea
and hence $v=6eJ$.  The velocity bound for the spreading of correlations
is obtained by setting the bound on $C_{i,j}(t)$ [the right-hand-side of
Eq. (\ref{eq:cijbound})] to a constant value and solving
$r=v_{\rm LR}t$, which yields $v_{\rm LR}=2v=12eJ$.

\subsection{Bound for commuting Hamiltonians}
Motivated by applications to the Ising model studied in the
manuscript, here we derive a bound applicable to 1D Hamiltonians
\begin{equation}
H=\sum_{k<l}h_{kl},
\end{equation}
where $[h_{kl},h_{k^{\prime}l^{\prime}}]=0$ for any
$k,l,k^{\prime},l^{\prime}$.  As above, we are interested in
bounding the connected correlation function $C_{i,j}(t)$, and without loss of
generality we take $i<j$.  For convenience in what follows, we define
$h_{kk}=0$, and take
$h_{kj}=h_{jk}$ (even though only one of the two appears in the
Hamiltonian).

To compute $A_{i}(t)$, let us first define $H_{i}=\sum_{k}h_{ik}$
as the part of $H$ that (in general) does not commute with $A_{i}$, so that $A_{i}(t)=e^{iH_{i}t}A_{i}e^{-iH_{i}t}$.
We can further separate $H_{i}$ into two parts by choosing a site
index $k_0$ satisfying $i\leq k_0<j$ and writing
\begin{eqnarray}
H_{i}^{\prime} = \sum_{k\leq k_0}h_{ik},~~~~~H_{i}^{\prime\prime}=\sum_{k>k_0}h_{ik}.
\end{eqnarray}
As a result,
\begin{eqnarray}
A_{i}(t) & = & e^{iH_{i}^{\prime}t}e^{iH_{i}^{\prime\prime}t}A_{i}e^{-iH_{i}^{\prime\prime}t}e^{-iH_{i}^{\prime}t} \nonumber \\
 & = & e^{iH_{i}^{\prime}t}(A_{i}+\int_{0}^{t}d\tau[e^{iH_{i}^{\prime\prime}\tau}A_{i}e^{-iH_{i}^{\prime\prime}\tau},H_{i}^{\prime\prime}])e^{-iH_i^{\prime}t} \nonumber \\
 & \equiv & A_{i}^{\prime}(t)+f_{i}(t),
\end{eqnarray}
where $A_{i}^{\prime}(t)=e^{iH_{i}^{\prime}t}A_{i}e^{-iH_{i}^{\prime}t}$ and
\ba
\lVert f_{i}(t)\rVert \le2t\lVert A_{i}\rVert \lVert H_{i}^{\prime\prime}\rVert.
\ea
Similarly, we can define
\begin{eqnarray}
H_{j}^{\prime}&=&\sum_{k>k_0}h_{jk},~~~~~H_{j}^{\prime\prime}=\sum_{k\le k_0}h_{jk},
\end{eqnarray}
and $A_{j}^{\prime}(t)=e^{iH_{j}^{\prime}t}A_{j}e^{-iH_{j}^{\prime}t}$,
such that $A_{j}(t)=A_{j}^{\prime}(t)+f_{j}(t)$ and
\ba
\lVert f_{j}(t)\rVert \le2t\lVert A_{j}\rVert \lVert H_{j}^{\prime\prime}\rVert.
\ea

In terms of these newly defined quantities, we can write
\begin{eqnarray}
C_{i,j}(t) & = & \langle A_{i}^{\prime}(t)A_{j}^{\prime}(t)\rangle_{c}+\langle f_{i}(t)A_{j}(t)\rangle_{c}+\langle A_{i}^{\prime}(t)f_{j}(t)\rangle_{c}\nonumber,
\end{eqnarray}
where we note that the second term contains $A_j(t)$ [rather than
$A^{\prime}_j(t)$].  By inspection, $\langle
A_{i}^{\prime}(t)A_{j}^{\prime}(t)\rangle_{c}=0$.  Using the bounds on
$\lVert f_{i}(t)\rVert$ and $\lVert f_{j}(t)\rVert$, together with the
inequality $\left|\langle AB\rangle_{c}\right|\le2\left\Vert A\right\Vert \left\Vert B\right\Vert $, we find
\ba
\left|C_{i,j}(t)\right|\le4t\lVert A_{i}\rVert \lVert A_{j}\rVert \left(\left\Vert H_{i}^{\prime\prime}\right\Vert +\lVert H_{j}^{\prime\prime}\rVert \right).
\ea
Noting that $|J_{kl}|=\left\Vert h_{kl}\right\Vert $, then
\begin{eqnarray}
\left|C_{i,j}(t)\right|&\le&4t\left\Vert A_{i}\right\Vert \left\Vert A_{j}\right\Vert \left(\sum_{k>k_{0}} |J_{ik}|
+\sum_{k\le k_{0}}|J_{jk}|\right).\nonumber 
\end{eqnarray}
One can optimize the value of $k_{0}$ to give the tightest bound.
For power law couplings $J_{kl}\approx|k-l|^{-\alpha}$ ($\alpha>0$) in
1D, choosing $k_{0}$ right in the middle of $i$ and $j$ will generally give
the tightest bound.

\subsection{Multi-hop processes are forbidden for commuting Hamiltonians}
Here we prove the claim that, given an initial product state evolving
under a commuting Hamiltonian, distant spins
can only become correlated if they are
either directly coupled or if they
share an intermediate spin to which they both couple; multi-hop
processes (e.g. site $A$ coupling to site $D$ through sites $B$ and
$C$) do not occur.

We consider the time evolution of the operators $A_i$ and
$A_{j}$, residing on sites $i$ and $j$ of the lattice.  As discussed
in the previous section, the time evolution of $A_{i}$ and $A_{j}$ can
be written as
\begin{equation}
A_{i}(t)=e^{iH_it}A_{i}e^{-iH_it}~~~{\rm and}~~~A_{j}(t)=e^{iH_jt}A_{j}e^{-iH_jt},
\end{equation}
where
\begin{equation}
H_{i}=\sum_{p}h_{ip}~~~{\rm and}~~~H_{j}=\sum_{q}h_{jq}.
\end{equation}

We can expand the time-evolution operator to obtain
\begin{eqnarray}
\label{eqn:Uoperator}
&&A_i(t)= A_i+it [ H_i,A_i] -\frac{t^2}{2!}\left[ H_i ,
  [H_i,A_i]\right]+\ldots\\
&=& A_i+it \sum_{p_1}\left[ h_{ip_1},A_i \right]-\frac{t^2}{2!} \sum_{p_1,p_2}\left[h_{ip_2},\left[ h_{ip_1},A_i \right]\right]  + \ldots \nonumber 
\end{eqnarray}
It is clear from Eq. (\ref{eqn:Uoperator}) that $A_i(t)$ is supported
on (i.e. can be written in terms of operators belonging to) site $i$ and any site $p$ for which $\lVert
h_{ip}\rVert\neq 0$; we denote the set of such points by $\Lambda_i$,
and define an equivalent set $\Lambda_j$ containing all sites supporting the
operator $A_j(t)$.  If $\lVert h_{ij}\rVert=0$ and there are no sites $p$ that \emph{simultaneously} satisfy
$\lVert h_{ip}\rVert\neq 0$ and $\lVert h_{jp}\rVert\neq 0$, then
$\Lambda_i\cap\Lambda_j=\emptyset$.  In this case, it is clear that an
initial product state must satisfy
$\langle A_{i}(t)A_j(t)\rangle=\langle A_{i}(t)\rangle\langle
A_j(t)\rangle$, and therefore any connected
correlation function $C_{i,j}(t)$ must vanish.

\subsection{Numeric solutions}

Because no analytic solution exists for the $XY$ model, exact long-time
dynamics (where the perturbative results derived above break down)
must be obtained by numerical solution of the Schr\"odinger equation.
The curves presented in Fig.\ 3(m-n) are calculated using the {\tt
  NDSolve} function in Mathematica.  With our experimental spin-spin
couplings $J_{ij}$ as inputs [see Eq.~(6)], we construct the full $XY$
Hamiltonian [Eq.~(2)] using sparse matrices. After evolving the initial
product state $\ket{\psi(0)}$ under the Hamiltonian $H_{XY}$ for a time $t$, we construct the desired correlation functions by calculating
\begin{eqnarray}
C_{i,j}(t)&=&\bra{\psi(t)} \sigma_i^z \sigma_j^z \ket{\psi(t)} \nonumber \\
&-&\bra{\psi(t)} \sigma_i^z \ket{\psi(t)}\bra{\psi(t)} \sigma_j^z \ket{\psi(t)}.
\end{eqnarray}
\

\noindent To numerically check the light-cone shape when $\alpha=1.19$ in a
system of 22 spins, we follow a similar procedure but use MATLAB to
calculate the time-evolved state $\ket{\psi(t)}$. The results of this
calculation are shown in Fig. \ref{fig:LargeSimulation}.  Note that faster-than-linear growth of the light-cone boundary persists in this larger system of 22 spins.

\SuppFigureOne

\subsection{Short-time perturbation theory for the $XY$ model}
Unlike in the Ising model, no exact analytic solution exists for the
$XY$ model (even in 1D, owing to the long-range couplings).  However, we
can nevertheless expand the time-evolution operator to low order and
thereby recover the dynamics at short times.  At
sufficiently long times, this perturbative expansion (carried out here to second order) becomes a poor
approximation.  This failure, which is observed in the experimental
dynamics (Figure 4 of the manuscript),
suggests that the growth of correlations at long distances is not
the result of direct spin-spin interactions; instead those
correlations originate from the repeated propagation
of information through intermediate spins.

We are interested in the time evolution of a connected correlation
function $C_{i,j}(t)=\langle A_{i}(t)A_{j}(t)\rangle_{c}$ of observables $A_{i}$ and $A_{j}$ located at different sites $i$
and $j$. To second order in time, we have $A_{i}(t)=
A_{i}+it[H,A_{i}]-\frac{t^{2}}{2!}[H,[H,A_{i}]]+\mathcal{O}(t^3)$,
which yields
\begin{eqnarray}
\label{eqn:AitAjtc}
\langle A_{i}(t)A_{j}(t)\rangle_{c} &=& \langle
A_{i}A_{j}\rangle_{c}\\
&+&it\left(\langle A_{i}[H,A_{j}]\rangle_{c}+\langle[H,A_{i}]A_{j}\rangle_{c}\right)\nonumber\\
 &-&\frac{t^{2}}{2}\left(\langle A_{i}[H,[H,A_{j}]]\rangle_{c}+\langle[H,[H,A_{i}]]A_{j}\rangle_{c}\right)\nonumber\\
&-&t^2\langle[H,A_{i}][H,A_{j}]\rangle_{c}+\mathcal{O}(t^3).\nonumber
\end{eqnarray}
Note that in Eq. (\ref{eqn:AitAjtc}) we assume the notation $\langle
A_{i}[H,A_{j}]\rangle_{c}=\langle A_{i}[H,A_{j}]\rangle-\langle
A_{i}\rangle\langle[H,A_{j}]\rangle$.

In the experiment, where $A_{i}$ corresponds to the Pauli spin operator
$\sigma_i^z$, the initial state is: (1) a product state
$\ket{\downarrow\cdots\downarrow}_z$, and (2) a simultaneous eigenstate of each $A_i$. As a result of (1), the
connected correlation at $t=0$ vanishes ($\langle
A_{i}A_{j}\rangle_{c}=0$).  As a result of (2), the second and third
lines in Eq. (\ref{eqn:AitAjtc}) vanish. Therefore we have,
\begin{equation}
\label{eqn:afterthreezeros}
\langle \sigma^z_{i}(t)\sigma^z_{j}(t)\rangle_{c}=-t^{2}\langle[H,\sigma^z_{i}][H,\sigma^z_{j}]\rangle_{c}+\mathcal{O}(t^3).
\end{equation}
For the $XY$ Hamiltonians we find
\begin{equation}
[H,\sigma^z_{i}]=-i\sum_{k\ne i}J_{ik}\sigma_{i}^y \sigma_{k}^x,
\end{equation}
and so
\begin{eqnarray}
\label{eqn:almostfinalAiAj}
\langle
&&\sigma^z_{i}(t)\sigma^z_{j}(t)\rangle_{c}=\\
& &t^{2}\!\!\!\sum_{k\ne
  i,l\neq j}\!\!\!J_{ik}J_{jl}\left(\langle \sigma_{i}^y \sigma_{k}^x \sigma_{j}^y \sigma_{l}^x \rangle-\langle \sigma_{i}^y \sigma_{k}^x \rangle\langle \sigma_{j}^y \sigma_{l}^x \rangle\right)+\mathcal{O}(t^3).\nonumber
\end{eqnarray}
Since the initial state is polarized along $z$, the only term that
has a nonzero expectation value on the right hand side of
Eq. (\ref{eqn:almostfinalAiAj}) is the one with $k=j$ and $l=i$. Therefore,
\begin{eqnarray}
\label{eqn:finalAiAj}
\langle \sigma^z_{i}(t)\sigma^z_{j}(t)\rangle_{c} & = & t^{2}J_{ij}^{2}\langle \sigma_{i}^y \sigma_{j}^x \sigma_{j}^y \sigma_{i}^x\rangle+\mathcal{O}(t^3)\nonumber \\
 & = & t^{2}J_{ij}^{2}\langle \sigma_{i}^z \sigma_{j}^z\rangle+\mathcal{O}(t^3) \nonumber \\
& = & (J_{ij}t)^{2}+\mathcal{O}(t^3),
\end{eqnarray}
which is the short-time result used in the manuscript.

\end{document}